\documentclass[iop]{emulateapj}
\usepackage{epsfig}

\newcommand{\starnine}{K2-29}
\newcommand{\planetnine}{K2-29\,b}
\newcommand{\starseven}{K2-30}
\newcommand{\planetseven}{K2-30\,b}

\newcommand{\teff}{$T_{\rm eff}$}

\newcommand{\feh}{[Fe/H]}

\newcommand{\attw}{{ATLAS12}}           

\slugcomment{Accepted for Publication in AJ}

\begin{document}

\shorttitle{Two New Hot Jupiters} 
\shortauthors{Johnson et al.}

\title{Two Hot Jupiters from K2 Campaign 4}

\author{Marshall C. Johnson\altaffilmark{1}, Davide Gandolfi\altaffilmark{2,3}, Malcolm Fridlund\altaffilmark{4,5,6}, Szilard Csizmadia\altaffilmark{7}, Michael Endl\altaffilmark{1}, \\ Juan Cabrera\altaffilmark{7}, William D. Cochran\altaffilmark{1}, Hans J. Deeg\altaffilmark{8,9}, Sascha Grziwa\altaffilmark{10}, Ivan Ram\'irez\altaffilmark{1}, Artie P. Hatzes\altaffilmark{11}, Philipp Eigm\"uller\altaffilmark{7}, Oscar Barrag\'an\altaffilmark{2}, Anders Erikson\altaffilmark{7}, Eike W. Guenther\altaffilmark{11}, Judith Korth\altaffilmark{10}, Teet Kuutma\altaffilmark{12}, David Nespral\altaffilmark{8,9}, Martin P\"atzold\altaffilmark{10}, Enric Palle\altaffilmark{8,9}, Jorge Prieto-Arranz\altaffilmark{8,9}, Heike Rauer\altaffilmark{7,13}, \\ and Joonas Saario\altaffilmark{12}}

\altaffiltext{1}{Department of Astronomy and McDonald Observatory, University of Texas at Austin, 2515 Speedway, Stop C1400, Austin, TX 78712, USA; mjohnson@astro.as.utexas.edu}
\altaffiltext{2}{Dipartimento di Fisica, Universit\'a di Torino, via P. Giuria 1, 10125 Torino, Italy}
\altaffiltext{3}{Landessternwarte K\"onigstuhl, Zentrum f\"ur Astronomie der Universit\"at Heidelberg, K\"onigstuhl 12, 69117 Heidelberg, Germany}
\altaffiltext{4}{Max-Planck-Institut f\"ur Astronomie, K\"onigstuhl 17, 69117 Heidelberg, Germany}
\altaffiltext{5}{Leiden Observatory, University of Leiden, PO Box 9513, 2300 RA, Leiden, The Netherlands}
\altaffiltext{6}{Department of Earth and Space Sciences, Chalmers University of Technology, Onsala Space Observatory, 439 92 Onsala, Sweden}
\altaffiltext{7}{Institute of Planetary Research, German Aerospace Center, Rutherfordstrasse 2, 12489 Berlin, Germany}
\altaffiltext{8}{Instituto de Astrof\'isica de Canarias, 38205 La Laguna, Tenerife, Spain}
\altaffiltext{9}{Departamento de Astrof\'isica, Universidad de La Laguna, 38206 La Laguna, Spain}
\altaffiltext{10}{Rheinisches Institut f\"ur Umweltforschung an der Universit\"at zu K\"oln, Aachener Strasse 209, 50931 K\"oln, Germany}
\altaffiltext{11}{Th\"uringer Landessternwarte Tautenburg, Sternwarte 5, D-07778 Tautenberg, Germany}
\altaffiltext{12}{Nordic Optical Telescope, Apartado 474, 38700 Santa Cruz de La Palma, Spain}
\altaffiltext{13}{Center for Astronomy and Astrophysics, TU Berlin, Hardenbergstr. 36, 10623 Berlin, Germany}

\begin{abstract}

We confirm the planetary nature of two transiting hot Jupiters discovered by the {\it Kepler}~\ spacecraft's K2 extended mission in its Campaign 4, using precise radial velocity measurements from FIES@NOT, HARPS-N@TNG, and the coud\'e spectrograph on the McDonald Observatory 2.7 m telescope. K2-29 b (EPIC 211089792\,b) transits a K1V star with a period of $3.2589263\pm0.0000015$ days; its orbit is slightly eccentric ($e=0.084_{-0.023}^{+0.032}$). It has a radius of $R_P=1.000_{-0.067}^{+0.071}$ $R_J$ and a mass of $M_P=0.613_{-0.026}^{+0.027}$ $M_J$. Its host star exhibits significant rotational variability, and we measure a rotation period of $P_{\mathrm{rot}}=10.777 \pm 0.031$ days. K2-30 b (EPIC 210957318\,b) transits a G6V star with a period of $4.098503\pm0.000011$ days. It has a radius of $R_P=1.039_{-0.051}^{+0.050}$ $R_J$ and a mass of $M_P=0.579_{-0.027}^{+0.028}$ $M_J$. The star has a low metallicity for a hot Jupiter host, $[\mathrm{Fe}/\mathrm{H}]=-0.15 \pm 0.05$.
\end{abstract}


\section{Introduction}

More than two decades have now elapsed since the discovery of the first exoplanet orbiting a main sequence star \citep{MayorQueloz95}. Over that time tremendous technological advances have been made, allowing the detection of Earth-size and approximately Earth-mass planets with both transits \citep[e.g.,][]{JontofHutter15} or radial velocities \citep[RVs; e.g.,][]{Wright15}. Nonetheless, hot Jupiters remain the easiest planets to detect with both techniques, due to their large radii and masses and short orbital periods. The Exoplanet Orbit Database\footnote{http://exoplanets.org/} \citep{Han14} lists more than 200 known hot Jupiters ($P<10$ days, $0.3 M_J < M_P < 13 M_J$) as of January 2016. Expanding this sample will allow ever more fine-grained explorations of correlations among parameters for this class of planets (mass, radius, orbital period, temperature, stellar properties, etc.), as well as the discovery of rare systems. A recent example of this latter class is the WASP-47 system, which is unique in having two small planets orbiting in close proximity to the hot Jupiter \citep{Becker15}.

An expanded population of known hot Jupiters, and the statistical explorations and further observations that this will enable, can help to answer a number of unsolved problems regarding hot Jupiters. One such outstanding problem involves the origins of hot Jupiters. Ever since the discovery of the first hot Jupiter, 51 Peg b, it has been recognized that forming these planets {\it in situ} would be very difficult \citep{MayorQueloz95}. A wide variety of mechanisms have been proposed to bring hot Jupiters in from their formation locations outside the snow line to where they are observed today \citep[e.g.,][]{Lin96,RasioFord96,FabryckyTremaine07}, and a great deal of work has been devoted to determining which mechanisms actually produce hot Jupiters \citep[e.g.,][]{Naoz12,DawsonMurrayClay13}. Despite this, the contributions of different migration mechanisms to the hot Jupiter population, and even which is the dominant migration mechanism, is still a matter of debate. There have even been recent suggestions that some hot and warm Jupiters (i.e., Jovian planets with periods of greater than approximately 10 days but interior to the habitable zone--exact values depend upon the stellar parameters) might form {\it in situ} rather than migrating from further out \citep{Batygin15,Huang16}. The expansion of the population of hot Jupiters, enabling more follow-up observations and the discovery of rare objects like WASP-47, can help to solve this problem.

The most prolific planet hunter to date is the {\it Kepler} mission \citep{Borucki10}, which has produced thousands of planet candidates \citep{Coughlin15} and hundreds of validated or confirmed planets \citep[e.g.][]{Rowe14}. After the failure of a second reaction wheel, however, the {\it Kepler} spacecraft was no longer able to keep pointing at the original {\it Kepler} field, and was repurposed for the K2 extended mission \citep{Howell14}. During this extended mission it is surveying a series of fields around the ecliptic; each campaign (observations of a specific field) has a duration of $\sim75-80$ days. K2 has already produced hundreds of planet candidates \citep{Vanderburg15}, but, like any other transit survey, further observations are typically needed to confirm or validate planet candidates as {\it bona fide} planets. This is particularly true for giant planet candidates, as this population suffers from high false positive rates \citep{Fressin13,Santerne15}. Conversely, however, giant planet candidates are the easiest population to confirm with RV measurements of the stellar reflex motion due to the large RV semi-amplitude variations induced by the planets.

Here we present K2 photometry for two late-type dwarf stars, EPIC 211089792 (\starnine) and EPIC 210957318 (\starseven), for which we identified periodic transit signals, and our follow-up spectroscopic observations. These have allowed us to confirm both transiting objects as {\it bona fide} hot Jupiters, and to measure the stellar and planetary parameters.

\section{K2 Photometry}

Observations for K2 Campaign 4 began on 2015 February 7 UT and lasted until 2015 April 23 UT\footnote{http://keplerscience.arc.nasa.gov/k2-fields.html}; during these observations the boresight of the {\it Kepler} spacecraft was pointed at coordinates of $\alpha=03^{\mathrm{h}}56^{\mathrm{m}}18^{\mathrm{s}}$, $\delta=+18^{\circ}39'38''$. A total of 15,847 long cadence (30 minute integration time) and 122 short cadence (1 minute integration time) targets were observed.

We utilized two different methods to produce light curves for all 15,969 targets from the K2 pixel data. The first technique followed the methodology outlined in \cite{Grziwa15}. The K2 target pixel files were analyzed for stellar targets and a mask for each target was calculated. After light curve extraction, disturbances produced by the drift of the telescope were corrected by calculating the rotation of the telescope's CCD. This drift is caused by the fact that the {\it Kepler} spacecraft is now operating on only two reaction wheels, and is using the combination of carefully balanced solar radiation pressure and periodic thruster firings for stabilization about the third axis; this results in a periodic rotation of the spacecraft about the axis of the telescope \citep{Howell14}. In the second method, the photometric time-series data extraction was based on circular apertures. For each target, we selected an optimal aperture size to minimize the noise in the light curve and estimated the background by calculating the median value of the stamp after excluding all bright pixels which might belong to potential sources. The resulting light curves were decorrelated using the movement of the centroid as described in \cite{VanderburgJohnson14}. We also analyzed the light curves produced using their methodology, which are publicly available\footnote{https://www.cfa.harvard.edu/~avanderb/allk2c4obs.html}. We used aperture sizes of approximately 60 pixels to produce our own lightcurves, while the \cite{VanderburgJohnson14} lightcurves utilize aperture sizes of 29 and 18 pixels for \starnine\, and \starseven, respectively.

After extraction of the light curves, we searched for transit signals using the DST algorithm \citep{Cabrera12} and the EXOTRANS pipeline \citep{Grziwa12}. DST and EXOTRANS have been applied extensively to both {\it CoRoT} \citep{Carpano09,Cabrera09,Erikson12,Carone12,Cavarroc12} and {\it Kepler} \citep[][ Grziwa et al. 2016, submitted]{Cabrera14} data. All transit detection algorithms search for a pattern in the data and use a statistic to decide if a signal is present in the data or not. When compared to widely used algorithms like Box Least Squares \citep[BLS;][]{Kovacs02}, DST uses an optimized transit shape, with the same number of free parameters as BLS, and an optimized statistic for signal detection. EXOTRANS uses a combination of the wavelet based filter technique VARLET \citep{Grziwa15} and BLS. VARLET was developed to reduce stellar variability and discontinuities in light curves.

We identified periodic transit-like signals associated with two K2 Campaign 4 targets, EPIC 211089792 and EPIC 210957318; both signals were detected in the lightcurves produced by all three methods, and using both DST and EXOTRANS for transit detection. EPIC 211089792 was selected for K2 observations by program GO4007 (P.I. Winn), while EPIC 210957318 was also proposed by program GO4007 as well as programs GO4020 (P.I. Stello) and GO4060 (P.I. Coughlin). For brevity we will hereafter refer to these targets as \starnine\, and \starseven, respectively. \starnine\, was located on module 4, channel 10 of the {\it Kepler} focal plane, while \starseven\, was located on module 20, channel 71. Both targets passed all of the tests that we used to identify likely false positives (including lack of odd-even transit depth variations, absence of a deep secondary eclipse, and lack of large photometric variations in phase with the candidate orbital period), and so we proceeded to more detailed fitting of the light curve as well as reconnaissance spectroscopy and then RV observations (see \S\ref{specobs} for more on these latter points). We also searched for additional transit signals in the lightcurves of these stars, but none were found. The identifiers, magnitudes, colors, and proper motions of these stars are listed in Table~\ref{magstable}.

\starnine\, has a nearby star that is 4'' northeast of the target and $1.9\pm0.1$ mag fainter, and whose flux contaminates the measured light curve. Given that the aperture-mask that was used to extract the light curve is several times wider than the star-contaminator separation (1.0 pixels in the {\it Kepler} plate scale), we estimated that the fraction of the PSF from both target and contaminator that was included in the aperture mask is the same to within 20\%. We therefore used the contaminator-to-target brightness ratio as an estimate of the fraction of contaminating flux in the target light curve, obtaining $15\pm4\%$, and corrected the light curve accordingly. There are no sources close to \starseven\, that could contaminate its light curve.

In order to fit the K2 photometry and extract the transit parameters, we used the \texttt{EXOFAST} package \citep{Eastman13} to simultaneously fit the K2 photometry as extracted by \cite{VanderburgJohnson14} and our radial velocity observations. We modified the original code to account for RV data-sets from different spectrographs. Our results for \planetnine\, and \planetseven\, are discussed in \S\ref{planetnine} and \S\ref{planetseven}, respectively. 

\begin{deluxetable*}{lccr}
\tabletypesize{\scriptsize}
\tablecolumns{4}
\tablewidth{0pt}
\tablecaption{Stellar Identifiers, Magnitudes, and Colors \label{magstable}}
\tablehead{
\colhead{Parameter} &  \colhead{\starnine} & \colhead{\starseven} & \colhead{Source}
}

\startdata
\emph{Identifiers} & & & \\
\noalign{\smallskip}
EPIC  & 211089792 & 210957318  & EPIC \\
TYC  & 1818-1428-1 & \ldots & EPIC \\
UCAC  & 573-010529 & 562-007074 & EPIC \\
2MASS  & 04104086+2424061 & 03292204+2217577 & EPIC \\ 
WISE  & J041040.88+242405.9 & J032922.08+221757.7 & AllWISE \\
$\alpha$ (J2000.0)   & $04^{\mathrm{h}}10^{\mathrm{m}}40^{\mathrm{s}}.955$ & $03^{\mathrm{h}}29^{\mathrm{m}}22.071^{\mathrm{s}}$ & EPIC \\
$\delta$ (J2000.0)  & $+24^{\circ}24'07''.35$ & $+22^{\circ}17'57''.86$ & EPIC \\
\noalign{\smallskip}
\hline
\noalign{\smallskip}
\emph{Magnitudes}  & &  & \\
\noalign{\smallskip}
$B$  &  $13.16 \pm 0.36$ & $14.506 \pm 0.030$ & EPIC \\
$V$  &  $12.56 \pm 0.26$ & $13.530 \pm 0.040$ & EPIC \\
$g$  &  $12.928 \pm 0.020$ & $13.979 \pm 0.034$ & EPIC \\
$r$  &  $11.918 \pm 0.040$ & $13.189 \pm 0.040$ & EPIC \\
$i$  &  $12.908 \pm 0.900$ & $12.825 \pm 0.050$ & EPIC \\
$Kp$  &  12.914 & 13.171 & EPIC \\
$J$  &  $10.622 \pm 0.035$ & $11.632 \pm 0.019$ & 2MASS \\
$H$  &  $10.168 \pm 0.041$\tablenotemark{a} & $11.190 \pm 0.016$ & 2MASS \\
$K$  &  $10.062 \pm 0.034$\tablenotemark{a} & $11.088 \pm 0.020$ & 2MASS \\
$W1$  & $10.095 \pm 0.037$ & $11.016 \pm 0.023$ & AllWISE \\
$W2$  & $10.142 \pm 0.037$ & $11.058 \pm 0.021$ & AllWISE \\
$W3$  & $9.991 \pm 0.082$ & $11.067 \pm 0.161$ & AllWISE \\
$W4$  & $>7.549$ & $>9.003$ & AllWISE \\
\noalign{\smallskip}
\hline
\noalign{\smallskip}
\emph{Colors}  & &  & \\
\noalign{\smallskip}
$B-V$  &  $0.60 \pm 0.44$\tablenotemark{b} & $0.976 \pm 0.050$ & calculated \\
$J-K$  & $0.560 \pm 0.049$ & $0.544 \pm 0.028$ & calculated \\
\hline
\noalign{\smallskip}
\emph{Proper Motions}  & &  & \\
\noalign{\smallskip}
$\mu_{\alpha}\cos\delta$ (mas yr$^{-1}$) & $10.8\pm5.3$ & $25.9\pm2.3$ & UCAC4 \\
$\mu_{\delta}$ (mas yr$^{-1}$) & $-29.6\pm5.9$ & $-13.6\pm2.4$ & UCAC4
\enddata

\tablecomments{Stellar identifiers, magnitudes, and colors. Values of fields marked EPIC were taken from the Ecliptic Plane Input Catalog, available at http://archive.stsci.edu/k2/epic/search.php. Values marked 2MASS are from \cite{Skrutskie06}, those marked AllWISE are from \cite{Cutri14}, and those marked UCAC4 from \cite{Zacharias13}.}
\tablenotetext{a}{The 2MASS catalog notes that the $H$- and $K$-band magnitudes for \starnine\, are ``low quality results (upper limits or very poor photometry).''}
\tablenotetext{b}{The $B-V$ color that we calculated for \starnine\, is much bluer than expected for a K1V star; however, due to the poor-quality $B$ and $V$ photometry from EPIC the uncertainties on the color are very large. As noted in the text, the \texttt{isochrones} code predicts a color of $B-V=0.890_{-0.025}^{+0.026}$ (which is still consistent to within $1\sigma$ with that calculated from the photometry).}

\end{deluxetable*}

\section{High-Resolution Spectroscopy}
\label{specobs}

We obtained high-resolution spectroscopic observations of \starnine\, and \starseven\, using three different facilities. 

We used the Robert G. Tull coud\'e spectrograph \citep[TS23;][]{Tull95} on the 2.7\,m Harlan J. Smith Telescope at McDonald Observatory, Texas (USA), to obtain both reconnaissance spectroscopy (for initial vetting of stellar parameters) and RV observations. TS23 is a cross-dispersed slit-fed \'echelle spectrograph with a spectral resolving power of $R=60,000$. It has spectral coverage from 3750 \AA~to 10200~\AA, which is complete blueward of 5691~\AA. We used an I$_2$ cell for the RV observations. These observations occurred between November 2015 and January 2016.

We also obtained both reconnaissance spectroscopy and RV observations with the FIbre-fed \'Echelle Spectrograph \citep[FIES;][]{Frandsen99,Telting14} on the 2.56\,m Nordic Optical Telescope at the Observatorio del Roque de los Muchachos, La Palma (Spain). The observations were carried out between November 2015 and January 2016, as part of the observing programs OPTICON\,15B/064 and CAT\,15B/035, using the \emph{high-res} mode ($R=67,000$). This setup provides spectral coverage from 3640~\AA~to 7360~\AA. Following the same observing strategy as in \citet{Gandolfi15}, we traced the RV drift of the instrument by acquiring long-exposed (T$_\mathrm{exp}\sim35$~seconds) ThAr spectra immediately before and after each observation. We removed cosmic ray hits by combining 3 consecutive 1200 second sub-exposures per observation epoch.

Finally, we obtained RV observations with the HARPS-N spectrograph \citep[$R=115,000$, with wavelength coverage from 3830 to 6900 \AA;][]{Cosentino12} on the 3.58\,m Telescopio Nazionale Galileo, also at La Palma, between December 2015 and January 2016. These observations were part of the same observing programs as on FIES. Two consecutive exposures of 1800 seconds were acquired per observation epoch but were not combined.

We collected seven RV observations of \starnine\ with TS23, seven with FIES, and four with HARPS-N. For \starseven, we obtained four RVs each with FIES and HARPS-N. At $V=13.53$ mag, this star is too faint for TS23 iodine-cell RVs. All of our RVs, along with the bisector span measurements of the cross-correlation function (CCF) and the signal-to-noise ratio (SNR) per pixel at 5500 \AA, are listed in Table~\ref{spec_data}. We did not calculate CCF bisector spans for the TS23 data due to the presence of the iodine lines in the spectra. The RV measurements show no significant correlation with the CCF bisector spans, indicating that the observed Doppler shifts are induced by the orbital motion of the companions.

\begin{deluxetable*}{cccrccr}
\tabletypesize{\scriptsize}
\tablecolumns{6}
\tablewidth{0pt}
\tablecaption{Radial Velocity and Activity Measurements \label{spec_data}}
\tablehead{
\colhead{BJD$_\mathrm{TDB}$$-$2450000} & \colhead{RV (km s$^{-1}$)} & \colhead{$\sigma_{\mathrm{RV}}$ (km s$^{-1}$)}  & \colhead{BS (km s$^{-1}$)\tablenotemark{a}} & \colhead{Phase} & \colhead{SNR\tablenotemark{b}} & \colhead{Instrument}
}

\startdata
\noalign{\smallskip}
\multicolumn{7}{l}{\starnine} \\
\noalign{\smallskip}
7339.88236 & 28.1533 & 0.0258 & \ldots         &  0.52 & 31 & TS23 \\
7343.60425 & 32.7989 & 0.0119 & $-$0.0050      &  0.66 & 33 & FIES \\
7345.54925 & 32.6233 & 0.0290 & $-$0.0091      &  0.26 & 13 & FIES \\
7347.67108 & 32.7908 & 0.0214 & $-$0.0066      &  0.91 & 22 & FIES \\
7371.41031 & 32.9069 & 0.0122 & 0.0230         &  0.20 & 14 & HARPS-N \\
7371.43422 & 32.8797 & 0.0115 & 0.0389   &  0.20 & 14 & HARPS-N \\
7373.68822 & 28.2024 & 0.0455 & \ldots         &  0.90 & 18 & TS23 \\
7374.59449 & 28.0655 & 0.0389 & \ldots         &  0.17 & 20 & TS23 \\
7375.60615 & 28.1338 & 0.0310 & \ldots         &  0.48 & 24 & TS23 \\
7375.91120 & 28.1926 & 0.0384 & \ldots         &  0.58 & 22 & TS23 \\
7392.44628 & 32.7970 & 0.0121 & 0.0155         &  0.65 & 32 & FIES \\
7394.52107 & 32.6462 & 0.0121 & 0.0233         &  0.29 & 32 & FIES \\
7395.55871 & 32.7770 & 0.0123 & 0.0062         &  0.61 & 31 & FIES \\
7399.34590 & 33.0677 & 0.0080 & 0.0094         &  0.77 & 19 & HARPS-N \\
7399.36609 & 33.0736 & 0.0071 & 0.0187         &  0.77 & 20 & HARPS-N \\
7400.64452 & 28.1062 & 0.0348 & \ldots         &  0.17 & 20 & TS23 \\
7402.66084 & 28.3081 & 0.0411 & \ldots         &  0.79 & 20 & TS23 \\
7418.36906 & 32.7771 & 0.0240 & 0.0481    &  0.61   & 19 & FIES \\
\noalign{\smallskip}
\hline
\noalign{\smallskip}
\multicolumn{7}{l}{\starseven} \\
\noalign{\smallskip}
7343.46426 & 35.3591 & 0.0116 & $-$0.0052 &  0.23 & 27 & FIES \\
7345.65414 & 35.5187 & 0.0651 & $-$0.0702 &  0.77 &  5 & FIES \\
7370.63631 & 35.7079 & 0.0148 & $-$0.0280 &  0.86 & 10 & HARPS-N \\
7372.47940 & 35.5156 & 0.0296 & $-$0.0521 &  0.31 &  6 & HARPS-N \\
7394.46817 & 35.4922 & 0.0194 & $-$0.0275 &  0.68 & 21 & FIES \\
7395.41003 & 35.4697 & 0.0153 & $-$0.0107 &  0.91 & 24 & FIES \\
7399.43817 & 35.6695 & 0.0090 & $-$0.0356 &  0.89 & 15 & HARPS-N \\
7399.45999 & 35.6698 & 0.0094 & $-$0.0222 &  0.90 & 14 & HARPS-N 
\enddata

\tablecomments{Radial velocities and CCF bisector span measurements for our spectroscopic observations. The TS23 RVs are differential, not absolute, resulting in a very different value of the RV. The RV offset between the TS23 and FIES data is $\Delta\mathrm{RV}_{\mathrm{TS23-FIES}}=-4.5646\pm0.0031$ km s$^{-1}$.}
\tablenotetext{a}{For the CCF bisector spans we assumed an uncertainty twice that of the RV from the corresponding spectrum. We did not calculate bisector spans for the TS23 data due to the presence of iodine lines in the spectra.}
\tablenotetext{b}{Signal-to-noise ratio per pixel, measured at 5500 \AA.}

\end{deluxetable*}

In addition to measuring the RVs, we also used our high-resolution spectra to derive stellar parameters for our target stars. We stacked all of our FIES spectra for each target, and then analyzed these data adopting two independent procedures. The first method used a grid of theoretical models from \citet{Castelli04}, \citet{Coelho05}, and \citet{Gustafsson08} to fit spectral features that are sensitive to different photospheric parameters. We adopted the calibration equations from \citet{Bruntt10} and \cite{Doyle14} to estimate the microturbulent ($v_\mathrm{mic}$) and macroturbulent ($v_\mathrm{mac}$) velocities. We simultaneously fitted the spectral profiles of several clean and unblended metal lines to estimate the projected rotational velocity ($v\sin i_{\star}$). 

The second method relied on the Spectroscopy Made Easy (\texttt{SME}) package \citep{ValentiPiskunov96}. We used \attw\, model grids for the derivation of the stellar parameters, and again estimated the micro- and macroturbulent velocities using the same method as above. 
We primarily used the wings of the Balmer lines (mostly H$\alpha$ and H$\beta$) to determine \teff, and the Mg\,{\sc i} $\lambda$5167, $\lambda$5173, and $\lambda$5184~\AA, the Ca\,{\sc i} $\lambda$6162 and $\lambda$6439~\AA, and the Na\,{\sc i}~D ($\lambda$5890 and $\lambda$5896~\AA) lines to estimate $\log g_*$. 
In order to verify the accuracy of this method, we analyzed a Solar spectrum from \cite{hinkle}. Following the discussion given in \cite{Barklem2002}, we found the errors quoted to be representative of what can currently be achieved when calculating synthetic spectra in order to fit observations.  
Our final adopted values for $T_{\mathrm{eff}}$, $\log g_{\star}$, $[\mathrm{Fe}/\mathrm{H}]$, and $v\sin i_{\star}$ are the weighted means of the values produced by the two methods.

We derived the stellar mass and radius using \texttt{EXOFAST}, utilizing our values for $T_{\mathrm{eff}}$, $\log g_{\star}$ and $[\mathrm{Fe}/\mathrm{H}]$, the transit-derived stellar mean density, and the relationship between these parameters and $M_{\star}$, $R_{\star}$ found by \cite{Torres10}. In order to measure the age and distance of our targets, we used the \texttt{isochrones} package \citep{Morton15} to derive these parameters from our values of $T_{\mathrm{eff}}$, $\log g_{\star}$, and $[\mathrm{Fe}/\mathrm{H}]$ plus the available stellar magnitudes (we used the value of $\log g_{\star}$ derived from global modeling of the system, including the stellar density derived from the transit light curve, rather than the value of $\log g_{\star}$ derived purely from spectroscopy). We note, however, that the isochrone age uncertainties we report in Table~\ref{parstable} are formal, i.e., roughly proportional to the $1\sigma$ errors of the stellar parameters used as input, which do not include an estimate of systematic uncertainties. Thus, these isochrone age error bars could be severely underestimated.
For \starnine\, we also used the relations presented by \cite{Barnes07} to calculate the gyrochronological age of the star, as described in the next section.

\section{\planetnine}
\label{planetnine}

\begin{figure*}
\plotone{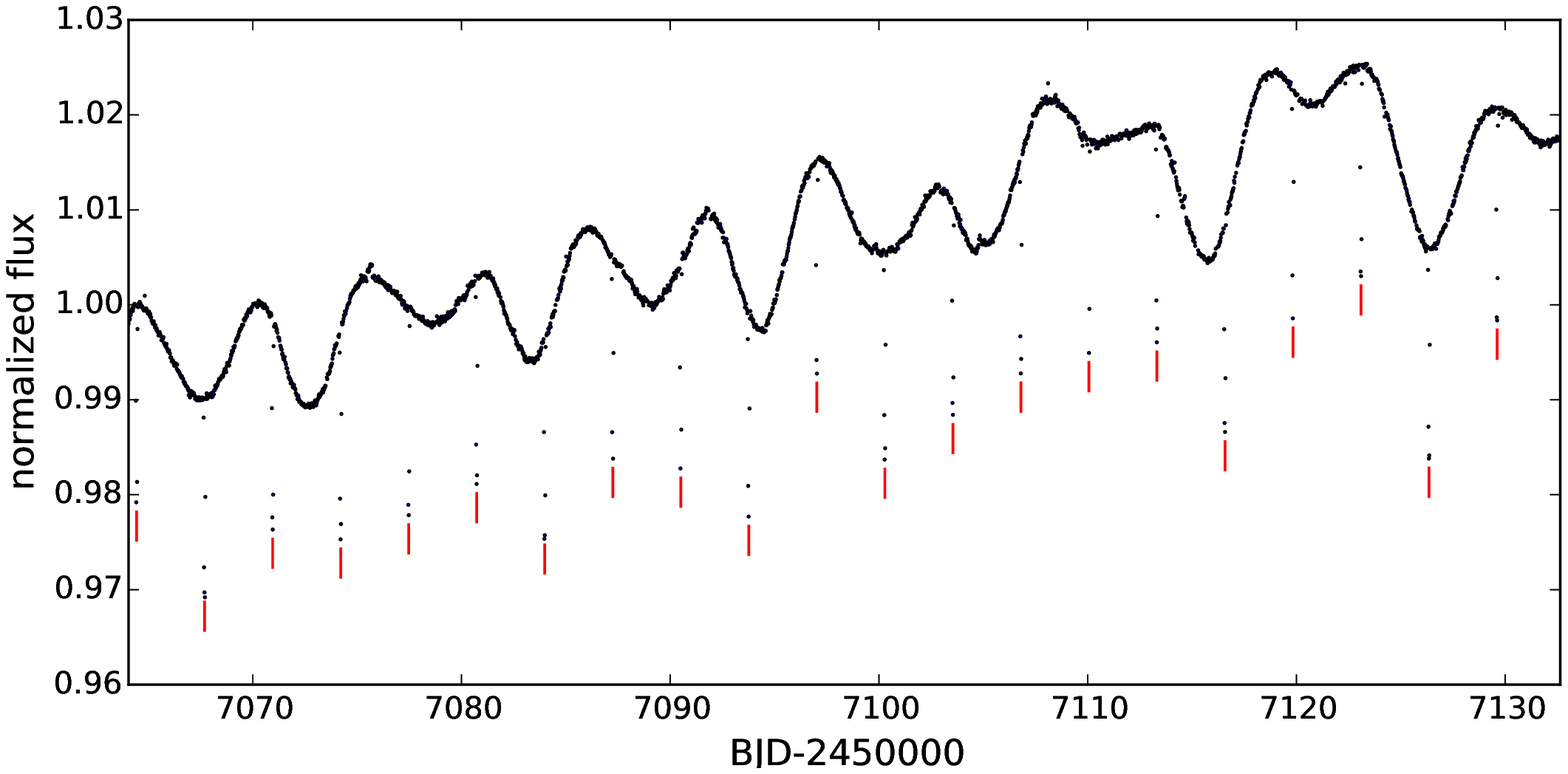}\\
\plottwo{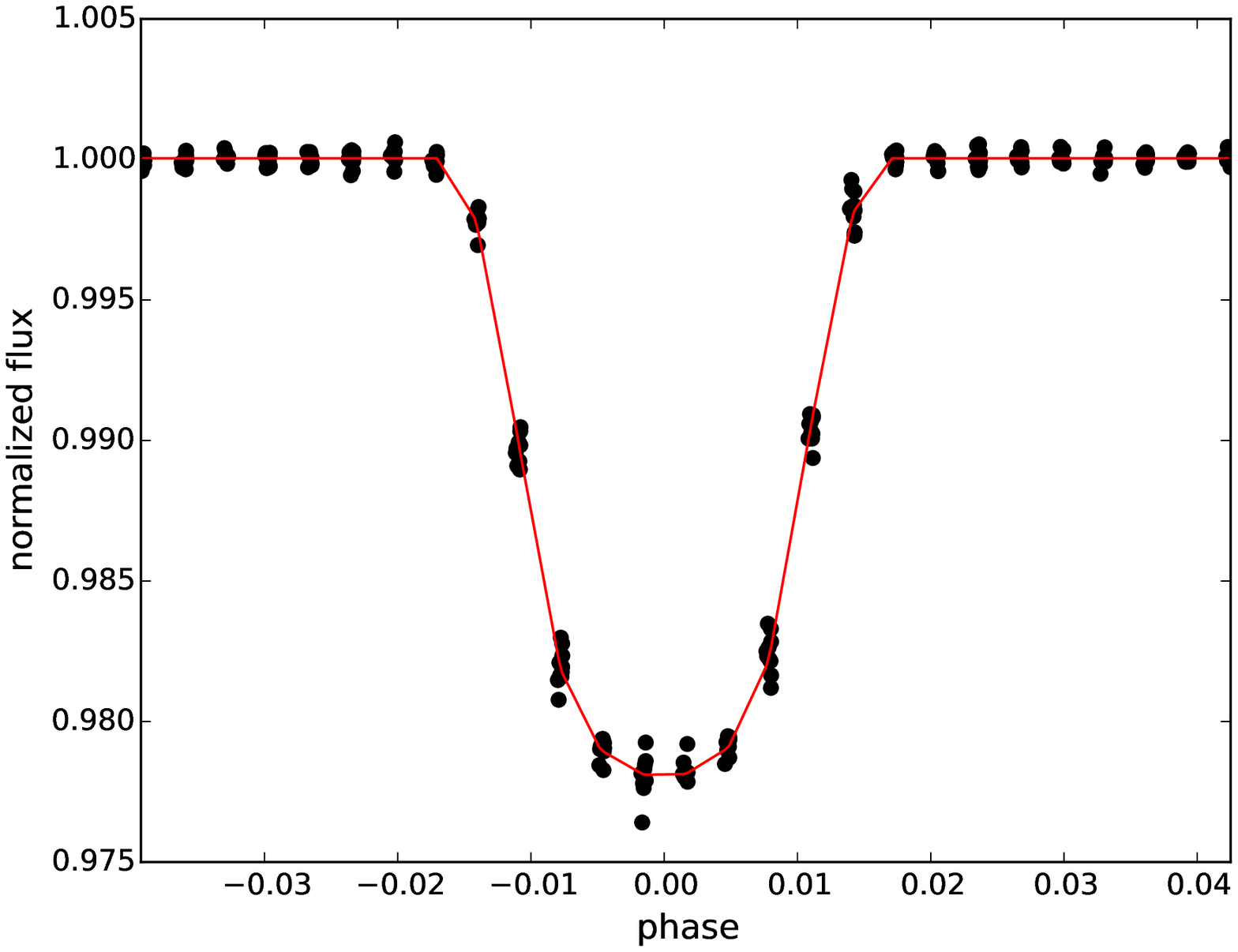}{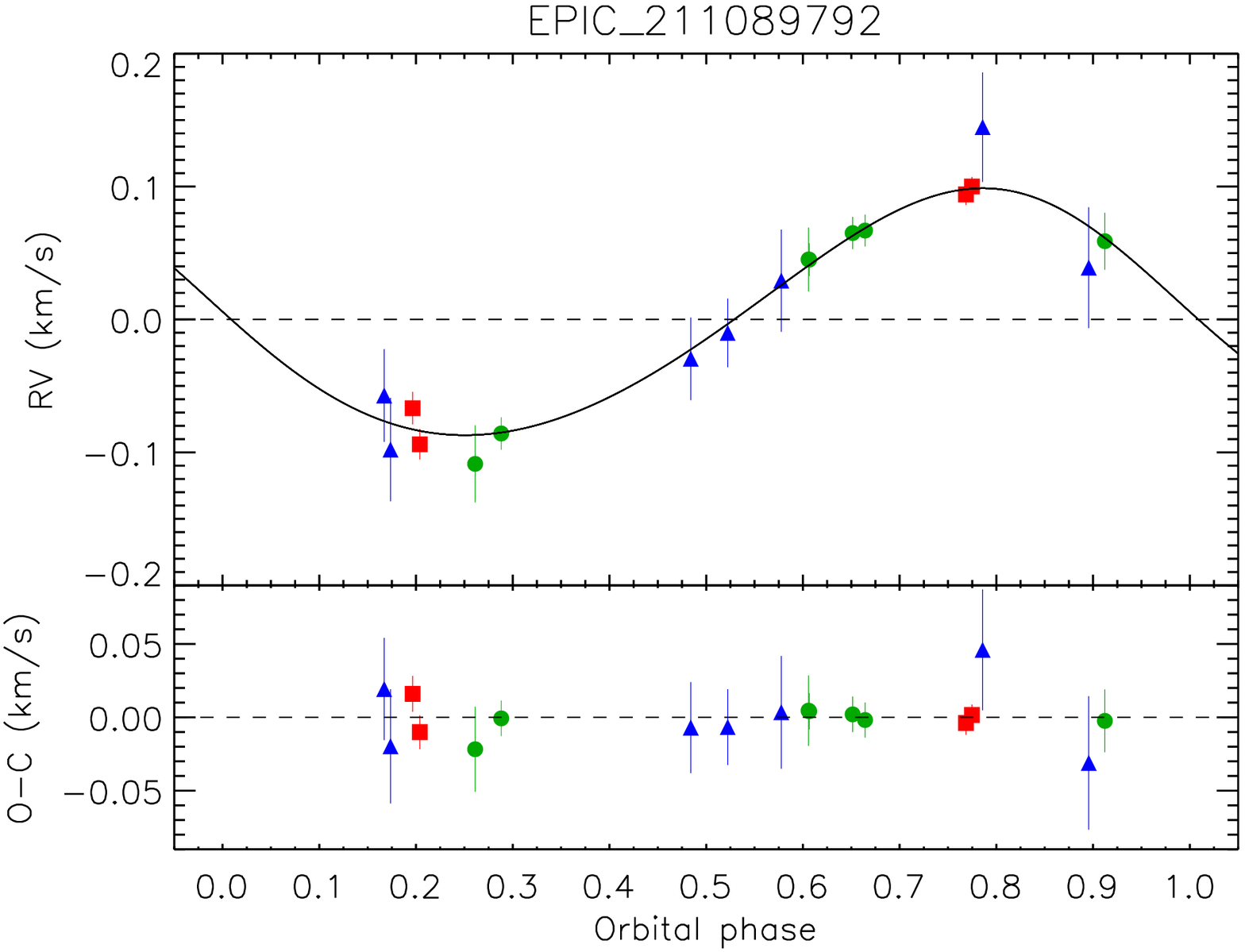}
\caption{Top: light curve for \starnine, produced using the pipeline of \cite{VanderburgJohnson14}. Times of transits are marked by vertical red bars. Note the large variability due to the stellar rotation. Bottom left: phase-folded light curve for \starnine. The best-fit model is overplotted in red. The clustering in the phase-folded data is due to the fact that the planetary orbital period ($3.2589265\pm0.0000015$ days) is very close to 156 {\it Kepler} 30 minute long cadence periods (3.25 days). Both the full and phase-folded lightcurves have been corrected for the contaminating flux of the nearby source, as described in the text. Bottom right: phase-folded RVs for \starnine, following the subtraction of the systemic velocity  and FIES-TS23 RV offset listed in Table~\ref{parstable}. HARPS-N data are shown with red squares, FIES with green circles, and McDonald with blue triangles.
\label{fig_9792}}
\end{figure*}

The full and phase-folded light curves and the RVs for \starnine\, are shown in Fig.~\ref{fig_9792} (clear outliers have been removed by visual inspection of the lightcurve). We used all 21 transits observed by K2 during Campaign 4 to obtain the best-fit parameters for the system. The parameters that we have measured and calculated for \starnine\, and \planetnine\, are listed in Table~\ref{parstable}. 

\begin{figure}
\plotone{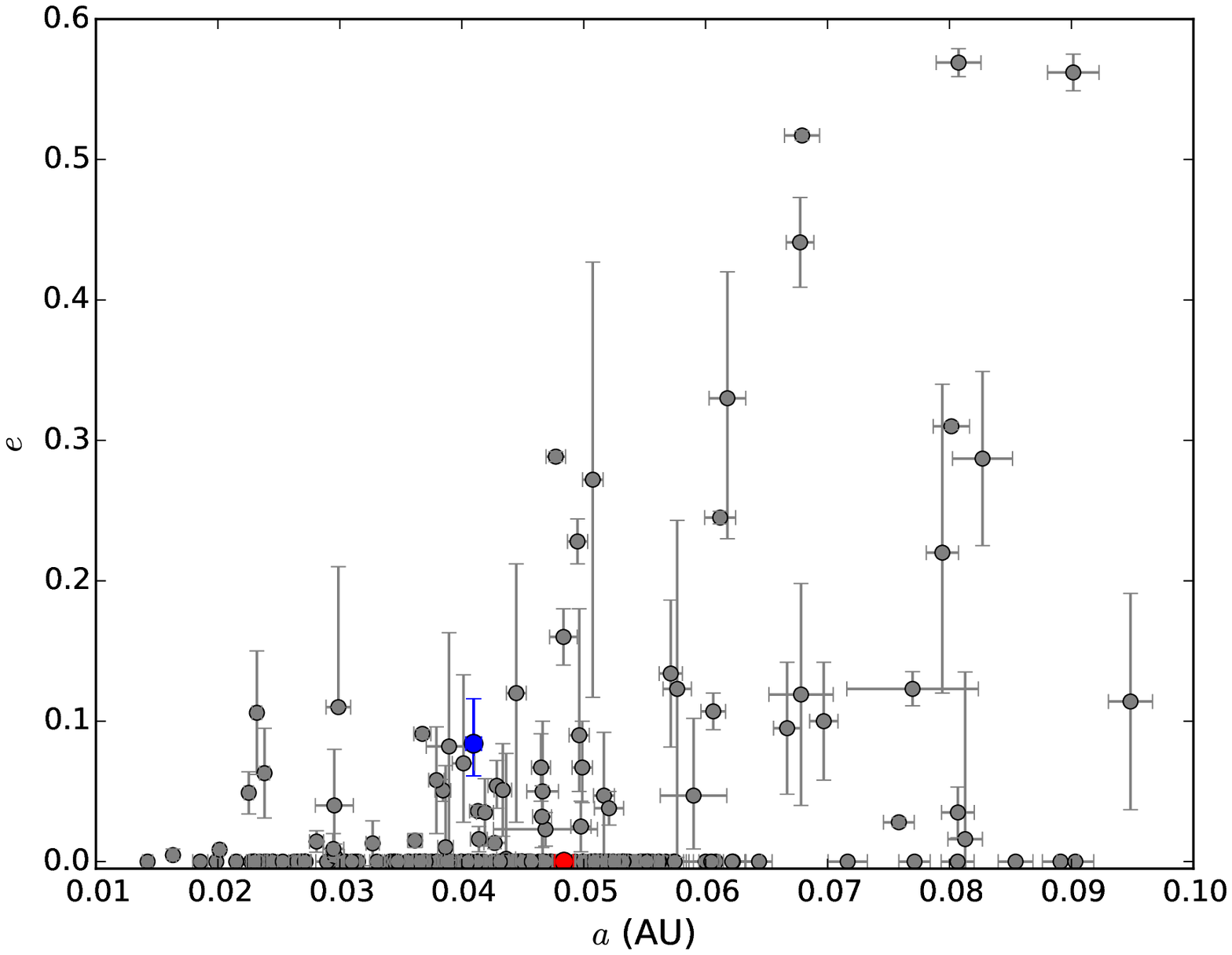}
\caption{Distribution of $e$ versus $a$ for known hot Jupiters ($P<10$ days, $0.3 M_J < M_P < 13 M_J$). \planetnine\, is marked in blue and \planetseven\, in red. Note that due to its circular orbit, \planetseven\, is shown at the bottom of the plot. The literature data are taken from the Exoplanets Orbit Database \citep{Han14} as of 2016 Jan. 8. \label{evsa}}
\end{figure}

\planetnine\, is a sub-Jupiter-mass ($0.613_{-0.026}^{+0.027} M_J$), Jupiter-radius ($1.000_{-0.067}^{+0.071} R_J$) planet orbiting a solar-metallicity ($[\mathrm{Fe}/\mathrm{H}]=0.00\pm0.05$) K1V star. It is perhaps most notable for its slightly eccentric orbit ($e=0.084_{-0.023}^{+0.032}$); its eccentricity is larger than that of most known hot Jupiters with similar orbital semi-major axes (Fig.~\ref{evsa}). We also fit a model with the eccentricity fixed to zero; we stress that imposing a circular orbit had a negligible effect (within 1$\sigma$) on the values of the other planetary parameters.

As can be seen from the K2 light curve (top panel of Fig~\ref{fig_9792}), \starnine\, exhibits significant rotational variability. The trough-to-peak amplitude is $\sim1-2\%$. The overall shape of the light curve changes significantly over the course of the K2 observations, indicating the presence of spot evolution and/or differential rotation. We analyzed the rotational variability using the auto-correlation function (ACF) of the light curve \citep[see e.g.,][]{McQuillan13}. Using this methodology, we found a rotation period of $P_{\mathrm{rot}}=10.777 \pm 0.031$ days. Periodogram analysis of the light curve gives a broadly consistent rotation period, although it also contains significant power at the first harmonic near 5.4 days. The ACF also displays a secondary correlation peak at $\sim$5.4 days that is likely caused by active regions at opposite longitudes of the star.

\subsection{Stellar Age}

Our values of the stellar age inferred from the isochrones ($2.6_{-1.1}^{+2.5}$ Gyr) and gyrochronology ($367\pm45$ Myr) are discrepant with each other. We used a stellar color of $B-V=0.890_{-0.025}^{+0.026}$ derived from isochrones using only our spectroscopically-derived stellar parameters in order to calculate an age using the gyrochronological relation of \cite{Barnes07}; we used this rather than the photometric value of $B-V=0.60\pm0.44$ due to the latter value's inconsistency with the spectral type of \starnine\, and its large uncertainty. Discrepancies between stellar ages inferred from isochrones and gyrochronology, however, are common for hot Jupiter host stars; these stars tend to rotate more quickly than expected given their isochrone ages and thus appear younger gyrochronologically \citep[e.g.,][]{Pont09,Maxted15}. This is thought to be due to interactions between the planet and the star--either through tidal spin-up of the star, or through the planet altering the star's magnetic field structure and thus reducing spin-down \citep{FerrazMello15}. The latter mechanism, however, is thought to be most efficient for F-type stars \citep{Lanza10}. We therefore adopted the isochrone age.

Despite the relatively old isochrone age, \starnine\, maintains sufficient spot coverage to result in rotational variability at a level of $\sim1-2\%$. We also detected chromospheric emission in the cores of the Ca~\textsc{ii} H and K lines; the spectra are too noisy at these wavelengths to obtain a quantitative measurement of the emission level, however. The presence of significant stellar activity suggests that either the star is relatively young (ages as young as 1.5 Gyr are still within $1\sigma$ of the isochrone age, using the formal uncertainties), or that tidal spin-up has allowed it to remain active despite its age.  

In order to further constrain the age we searched for Li absorption, which is an indicator of youth; GKM stars' primordial Li abundances are typically destroyed over the first few hundred Myr of their lives \citep[see e.g. the review of][]{Soderblom14}. We did not detect any Li absorption, and set a conservative upper limit of A(Li)$<1.3$ by fitting synthetic spectra to our stacked FIES spectrum. Our stacked HARPS-N spectrum is also consistent with this limit. Using the results of \cite{SestitoRandich05}, who measured Li abundances in young open clusters, we can easily rule out ages of $<250$ Myr for \starnine. Furthermore, we can likely rule out ages as great as $\sim600$ Myr; the more metal-rich clusters at this age studied by \cite{SestitoRandich05} show greater depletion, and for the clusters most similar in \feh\, to \starnine\, we would be able to detect the expected A(Li) level. \starnine\, therefore likely has an age of $>600$ Myr, which would rule out the gyrochronological value of $367\pm45$ Myr.

We also sought to determine whether \starnine\, could be a member of either of the young open clusters in the Campaign 4 field of view, the Hyades and Pleiades. We computed the space velocities of the star using the \texttt{gal\_uvw} IDL routine\footnote{http://idlastro.gsfc.nasa.gov/ftp/pro/astro/gal\_uvw.pro}, finding $U=-30.9$ km s$^{-1}$, $V=-18.0$ km s$^{-1}$, and $W=-20.4$ km s$^{-1}$. Using Eqn. 2 of \cite{Klutsch14} and the cluster $UVW$ values presented in that work, we estimated that \starnine\, has a 10\% probability of being a member of the Hyades Supercluster, but only a 0.01\% probability of belonging to the Pleiades Moving Group. Our calculated metallicity for \starnine, $[\mathrm{Fe}/\mathrm{H}]=0.00\pm0.05$, is also lower than that of the Hyades, which has $[\mathrm{Fe}/\mathrm{H}]\sim 0.15$ \citep{Perryman98}. We thus conclude that \starnine\, is unlikely to have originated as part of either the Pleiades or the Hyades.

\subsection{Tidal Evolution}

In order to quantitatively assess the implications of the eccentric planetary orbit and the tidal spin-up of the host star for the past evolution of the system, we calculated the tidal timescales of the system using the model of \cite{Leconte10}. We assumed values for the tidal quality factors $Q'$ of the star and planet of $10^6$ and $10^5$, respectively, and used our other measured parameters of the system. We found that the semi-major axis tidal interaction timescale is $[(da/dt)/a]^{-1}=1.77$ Gyr, and the eccentricity timescale is  $[(de/dt)/e]^{-1}=0.027$ Gyr. 

These results indicate that the planet is indeed donating angular momentum to the star, and as a result its semi-major axis is shrinking. The timescale for the star's rotational evolution with these tides, however, is $\sim60-70$ Gyr, indicating that the planet could {\emph not} have spun up the star through tides alone. The star's fast spin must therefore have been caused by another mechanism, for instance suppressed magnetic braking or the ingestion of another planetary body.

A value of the eccentricity tidal interaction timescale of 0.027 Gyr is much shorter than the age of the system, which has consequences for its past evolution. In general, eccentric orbits of hot Jupiters might be generated in two different manners: either the eccentricity is primordial, a relic of high-eccentricity migration that emplaced the planet on a short-period orbit, or the eccentricity is being excited by an external perturber. The short tidal eccentricity timescale suggests three possibilities for \planetnine: 1) it migrated recently via high-eccentricity migration; 2) its eccentricity is currently being excited by a perturber; or 3) we have underestimated the tidal quality factor of the planet and/or star, resulting in a much longer eccentricity timescale, in which case the planet could have migrated via high-eccentricity migration when the system was much younger. For instance, a value of $Q'_P=10^7$ would result in an eccentricity timescale of 2.24 Gyr, consistent with the system age. All of these scenarios are testable with further data.

If the orbital eccentricity of \planetnine\, is being excited by an external perturber, the presence of such a perturber could be detected through long-term radial velocity observations, or through transit timing variations (TTVs). As only long cadence K2 data are available for \starnine, and the $\sim80$ days of K2 observations are insufficient to be sensitive to TTVs due to a much longer-period perturber, we do not pursue this line of investigation. Future high-cadence transit observations could, however, be used to search for TTVs. 

High-eccentricity migration mechanisms include planet-planet scattering \citep[e.g.,][]{RasioFord96,Chatterjee08}, the Kozai-Lidov mechanism \citep[e.g.,][]{FabryckyTremaine07,Naoz12}, and coplanar high-eccentricity migration \citep{Petrovich15}. Planet-planet scattering and coplanar high-eccentricity migration are both expected to take place during the first tens to hundreds of Myr of a system's lifetime, whereas Kozai-Lidov cycles can cause migration resulting in the emplacement of hot Jupiters on short-period orbits even Gyrs after the system formed \citep{Petrovich15b}. If \planetnine\, migrated recently, then it likely did so via the Kozai-Lidov mechanism, but if the tidal quality factor is higher than expected and it migrated early in the system's history, it could have done so via any of these mechanisms. Further observations could help to distinguish between these mechanisms. All three mechanisms require the presence of an additional object in the system, which could be detected either through measurement of an RV trend, or high-resolution imaging to find a distant stellar companion which could induce Kozai-Lidov oscillations. Measurement of the spin-orbit misalignment of \planetnine\, would also be useful on this front; a misaligned orbit would point towards migration through either planet-planet scattering or Kozai-Lidov. 

Our measured rotational period and calculated stellar radius predict an equatorial rotational velocity of $v_{\mathrm{eq}}=3.51 \pm 0.22$ km s$^{-1}$. This deviates from the value of $v\sin i_{\star}=3.8\pm0.1$ km s$^{-1}$ that we measured from our spectra by only $\sim1.2\sigma$, 
suggesting that the star is viewed close to equator-on. This also suggests that the spin-orbit misalignment might be small, although large spin-orbit misalignments can exist even for stars with $i_{\star}\sim90^{\circ}$. Only direct measurement of the spin-orbit misalignment can solve this issue. 

\section{\planetseven}
\label{planetseven}

\begin{figure*}
\plotone{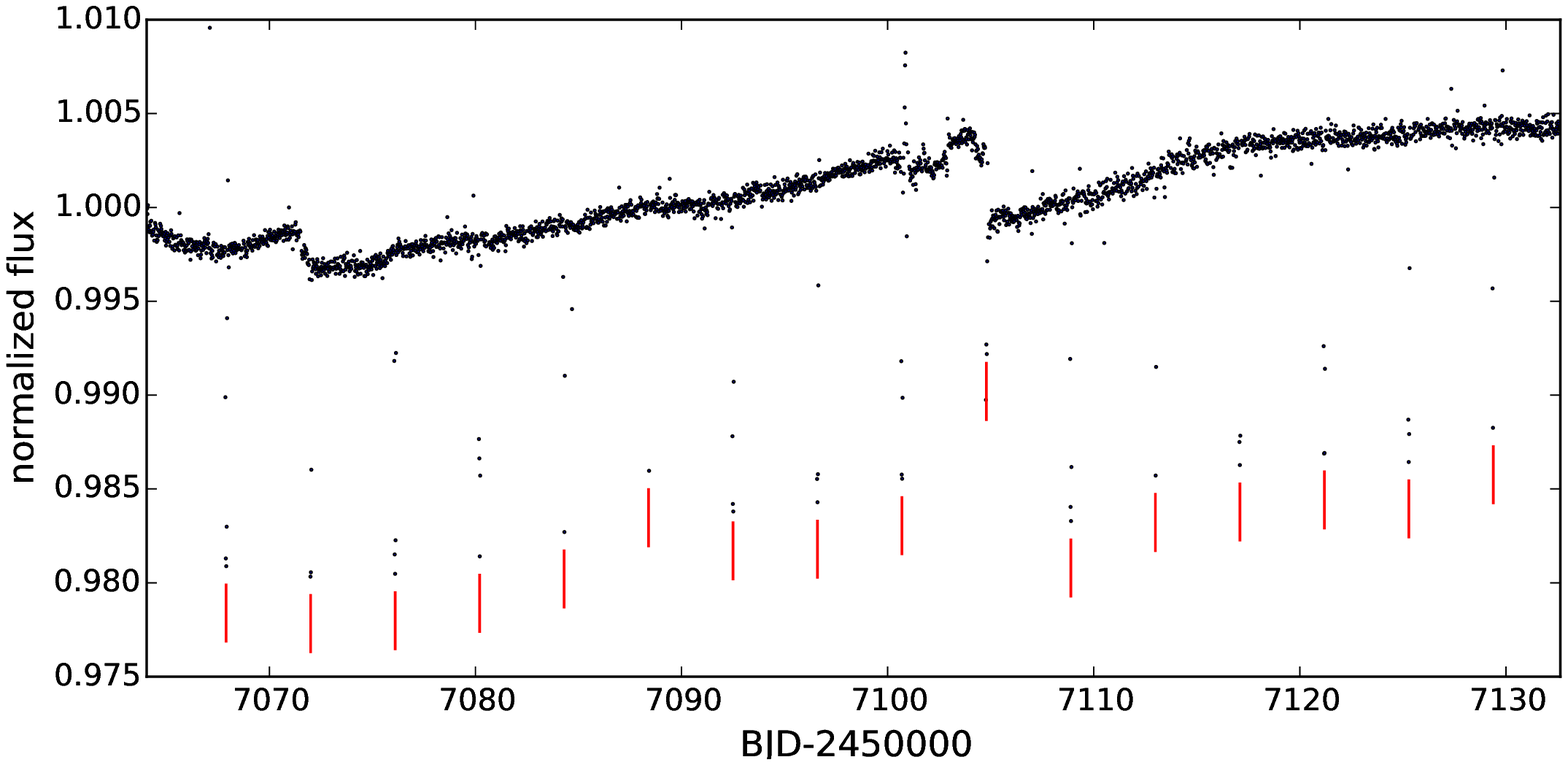}\\
\plottwo{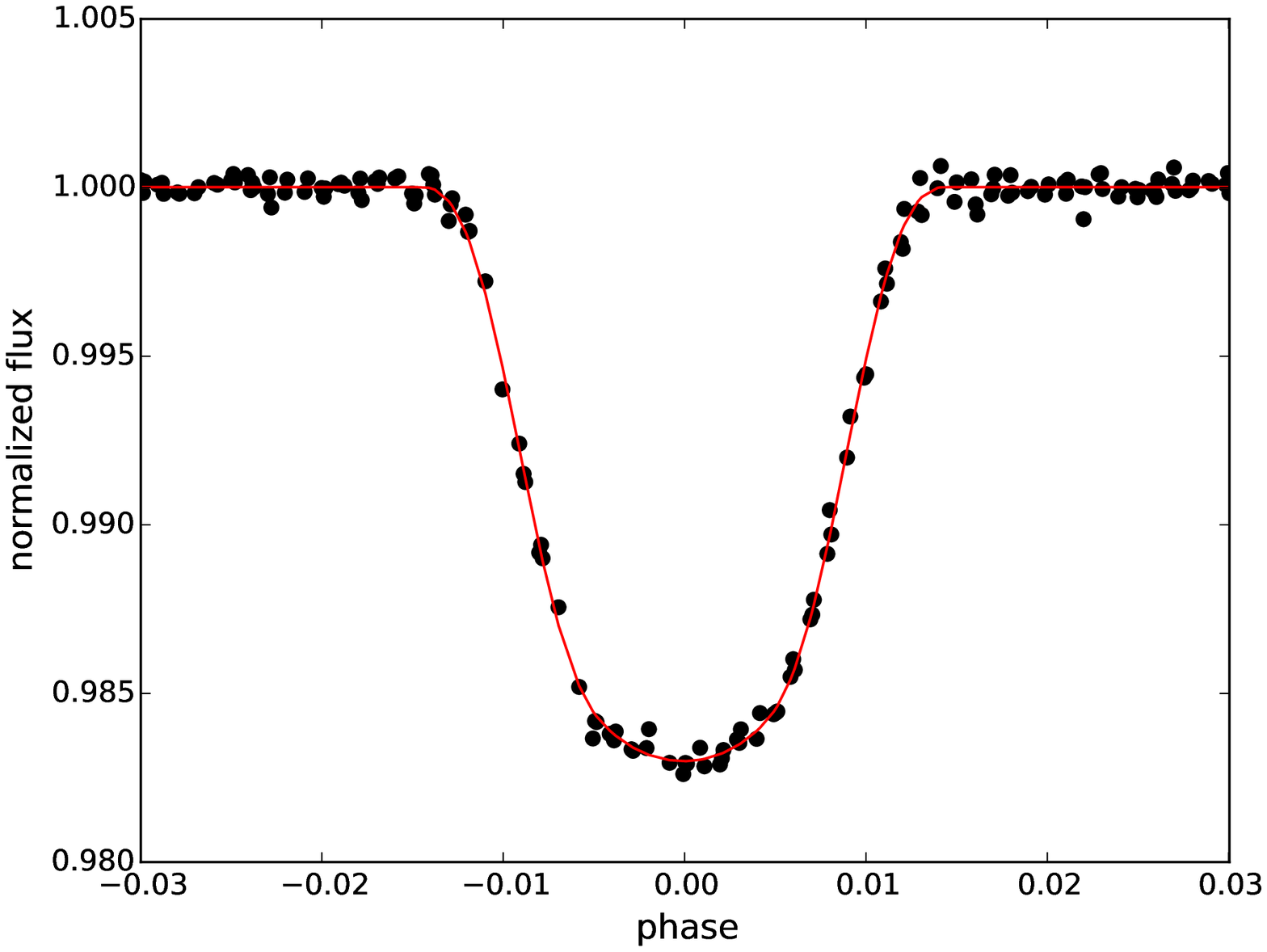}{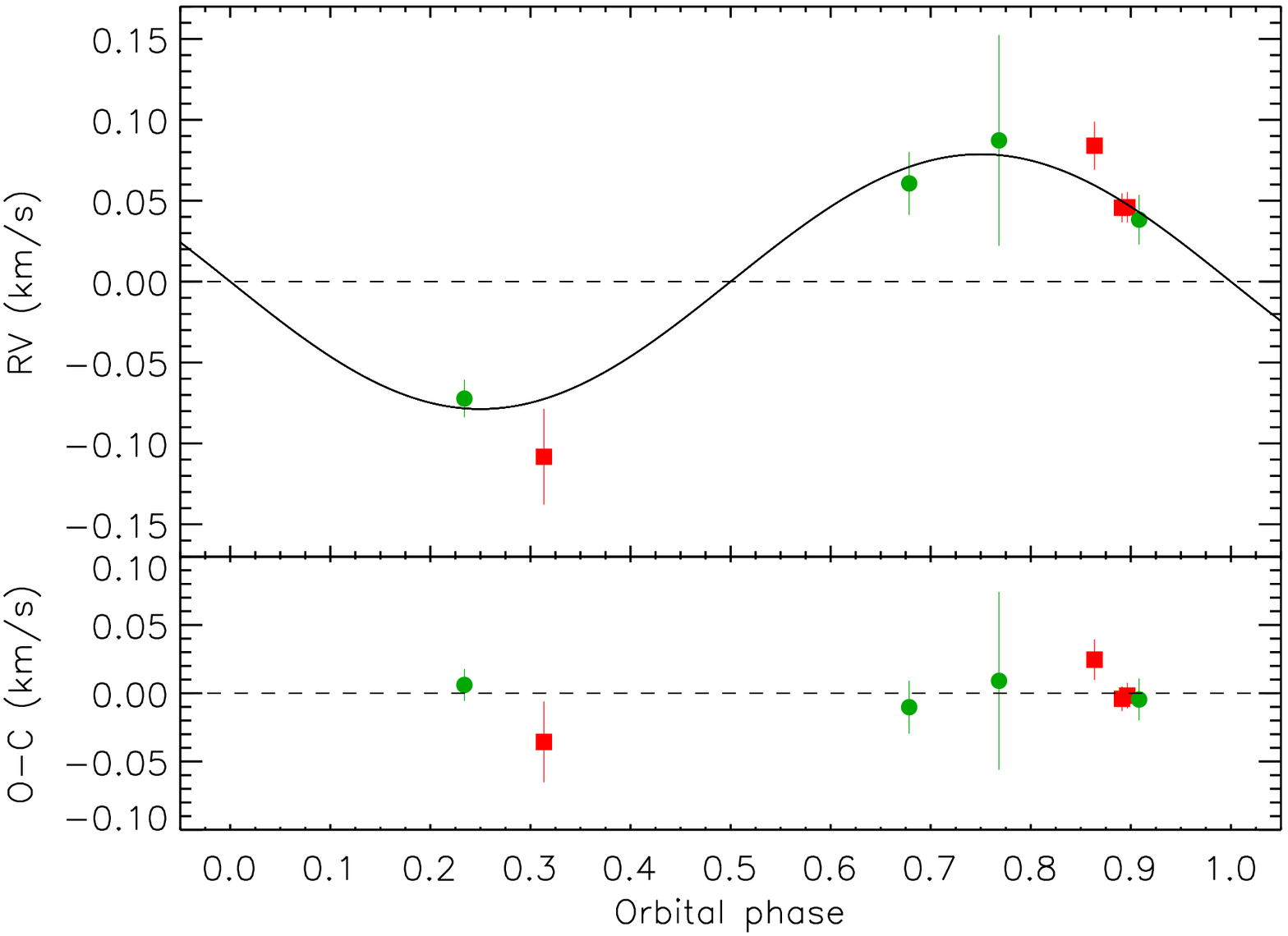}
\caption{Top: light curve for \starseven, produced using the pipeline of \cite{VanderburgJohnson14}. Times of transits are marked by vertical red bars. Several discontinuities in the light curve due to instrumental systematics are visible. Bottom left: phase-folded light curve for \starseven. The best-fit model is overplotted in red. We excluded the data from the 10$^\mathrm{th}$ transit from this figure (and from the fit) due to high noise and systematics. Bottom right: phase-folded RVs for \starseven, following the subtraction of the systemic velocity and FIES-TS23 RV offset listed in Table~\ref{parstable}. HARPS-N data are shown with red squares and FIES with green circles. 
\label{fig_7318}}
\end{figure*}

The full and phase-folded light curves and the RVs for \starseven\, are shown in Fig.~\ref{fig_7318}. A total of 16 transits were observed by K2 during Campaign 4. We removed clear outliers by visual inspection and excluded the 10$^\mathrm{th}$ transit from our analysis because of photometric discontinuities occurring immediately before and during the transit. We list the parameters for the star and planet in Table~\ref{parstable}.

\planetseven\, is a $0.579_{-0.027}^{+0.028} M_J$ planet with an orbital period of $4.098503\pm0.000011$ days. It has a radius of $1.039_{-0.051}^{+0.050} R_J$. We do not have sufficient RV data to constrain the planet's eccentricity, and so we fixed it to zero in our fits. 

The star has a slightly sub-solar metal content ($[\mathrm{Fe}/\mathrm{H}]=-0.15 \pm 0.05$). A metallicity this low is unusual for a hot Jupiter host, though by no means unprecedented; there are currently hot Jupiters known around stars with metallicities as low as $[\mathrm{Fe}/\mathrm{H}]=-0.6$ \citep[WASP-98;][]{Hellier14}. 

Using isochrones, we found an age of $3.9_{-1.9}^{+2.1}$ Gyr for \starseven. No significant rotational modulation is evident in the K2 light curve (top panel of Fig.~\ref{fig_7318}), and so we could not measure the rotation period and calculate the gyrochronological age for this star as we did for \starnine. We also calculated the space velocity of \starseven, finding $U=-43.9$ km s$^{-1}$, $V=-27.0$ km s$^{-1}$, and $W=-9.3$ km s$^{-1}$. Again using the methodology of \cite{Klutsch14}, we find that \starseven's kinematics give it a 7\% membership probability in the Hyades Supercluster, and an $8\times10^{-8}$\% membership probability in the Pleiades Moving Group. As is also the case for \starnine, \starseven's low metallicity is inconsistent with the bulk metallicity of the Hyades, and so we conclude that \starseven\, is also unlikely to have formed as part of either of these clusters.

\begin{deluxetable*}{llccr}
\tabletypesize{\scriptsize}
\tablecolumns{5}
\tablewidth{0pt}
\tablecaption{Stellar and Planetary Parameters \label{parstable}}
\tablehead{
\colhead{Parameter} & \colhead{Explanation} &  \colhead{\starnine} & \colhead{\starseven} & \colhead{Source}
}

\startdata
\multicolumn{5}{l}{\emph{Measured stellar parameters}} \\
\noalign{\smallskip}
SpT & spectral type\tablenotemark{a} & K1\,V & G6\,V & spectroscopy \\
$T_{\mathrm{eff}}$ & effective temperature (K) & $5222\pm40$ & $5425\pm40$ & spectroscopy \\
$\log g_*$ & surface gravity (cgs)\tablenotemark{b} & $4.54\pm0.07$ & $4.53\pm0.07$ & spectroscopy \\
$[\mathrm{Fe}/\mathrm{H}]$ & metallicity &  $0.00\pm0.05$ & $-0.15\pm0.05$ & spectroscopy \\
$v_{\mathrm{mic}}$ & microturbulent velocity\tablenotemark{c} (km s$^{-1}$) & $0.85\pm0.09$ & $0.91\pm0.09$ & spectroscopy \\
$v_{\mathrm{mac}}$ & macroturbulent velocity\tablenotemark{c} (km s$^{-1}$) & $2.33\pm0.52$ & $2.46\pm0.53$ & spectroscopy \\
$v\sin i_{\star}$ & projected rotational velocity (km s$^{-1}$) & $3.8\pm0.1$ & $1.4\pm0.3$ & spectroscopy \\
$\gamma_{\mathrm{FIES}}$ & systemic RV (km s$^{-1}$) & $32.7319\pm0.0029$ & $35.4314\pm0.0021$ & spectroscopy \\
$\gamma_{\mathrm{HARPS}}$ & systemic RV (km s$^{-1}$) & $32.9736\pm0.0034$ & $35.6238\pm0.0031$ & spectroscopy \\
$\Delta\mathrm{RV}_{\mathrm{TS23-FIES}}$ & TS23$-$FIES RV offset\tablenotemark{c} (km s$^{-1}$) & $-4.5686\pm0.0031$ & \ldots & spectroscopy \\
$P_{\mathrm{rot}}$ & rotation period (days) & $10.777\pm0.031$ & \ldots & photometry \\
\noalign{\smallskip}
$u_1$ & linear limb darkening coeff.\tablenotemark{e} & $0.484_{-0.040}^{+0.038}$ & $0.451\pm0.044$ & photometry \\
\noalign{\smallskip}
$u_2$ & quadratic limb darkening coeff.\tablenotemark{e} & $0.173_{-0.049}^{+0.050}$ & $0.231_{-0.048}^{+0.049}$ & photometry \\
\noalign{\smallskip}
$\rho_{\star}$ & density (g cm$^{-3}$) & $2.90_{-0.44}^{+0.51}$ & $2.11_{-0.19}^{+0.22}$ & photometry \\
\noalign{\smallskip}
\hline
\noalign{\smallskip}
\multicolumn{5}{l}{\emph{Derived stellar parameters}} \\
\noalign{\smallskip}
$M_{\star}$ & mass ($M_{\odot}$) & $0.864_{-0.039}^{+0.041}$ &  $0.900_{-0.042}^{+0.043}$ &  spectroscopy+$\rho_{\star}$+T10\tablenotemark{f} \\
\noalign{\smallskip}
$R_{\star}$ & radius ($R_{\odot}$) &  $0.748_{-0.042}^{+0.045}$ & $0.844\pm0.032$ & spectroscopy+$\rho_{\star}$+T10\tablenotemark{f} \\
\noalign{\smallskip}
$L_{\star}$ & luminosity ($L_{\odot}$) &  $0.374_{-0.043}^{+0.050}$ & $0.554_{-0.046}^{+0.048}$ & spectroscopy+$\rho_{\star}$+T10\tablenotemark{f} \\
\noalign{\smallskip}
$\log g_*$ & surface gravity (cgs)\tablenotemark{g} & $4.626_{-0.047}^{+0.046}$ & $4.540_{-0.027}^{+0.029}$ & spectroscopy+$\rho_{\star}$+T10\tablenotemark{f}\\
\noalign{\smallskip}
age & age (Gyr)\tablenotemark{h}  & $2.6_{-1.1}^{+2.5}$ & $3.9_{-1.9}^{+2.1}$ & isochrones \\
\noalign{\smallskip}
d & distance (pc) & $167.1_{-4.0}^{+3.8}$ & $278.0\pm7.6$ & isochrones \\
\noalign{\smallskip}
\hline
\noalign{\smallskip}
\multicolumn{5}{l}{\emph{Measured planetary parameters}} \\
\noalign{\smallskip}
$P_\mathrm{orb}$ & orbital period (days) &  $3.2589263\pm0.0000015$ &  $4.098503\pm0.000011$ & photometry \\
\noalign{\smallskip}
$T_0$ & transit epoch (BJD$_\mathrm{TDB}-$2450000) & $7383.80546\pm0.00013$ & $7063.80714\pm0.00010$ & photometry \\
\noalign{\smallskip}
$\tau_{14}$ & transit duration (days) & $0.08875\pm0.00100$ & $0.09670\pm0.00100$ & photometry \\
\noalign{\smallskip}
$\tau_{12}=\tau_{34}$ & ingress/egress duration (days) &  $0.0126\pm0.0015$ & $0.0181_{-0.0015}^{+0.0016}$ & photometry \\
\noalign{\smallskip}
$a/R_{\star}$ & scaled semi-major axis &  $11.77_{-0.62}^{+0.65}$ &  $12.34_{-0.38}^{+0.42}$ & photometry \\
\noalign{\smallskip}
$R_p/R_{\star}$ & radius ratio & $0.1373_{-0.0026}^{+0.0024}$ &  $0.1266_{-0.0016}^{+0.0015}$ & photometry \\
\noalign{\smallskip}
$\delta$ & transit depth (\%) & $1.884_{-0.070}^{+0.066}$ & $1.603_{-0.041}^{+0.037}$ & photometry \\
\noalign{\smallskip}
$b$ & impact parameter &  $0.452_{-0.15}^{+0.096}$ & $0.662_{-0.036}^{+0.030}$ & photometry \\
\noalign{\smallskip}
$i_p$ & orbital inclination ($^{\circ}$) &  $87.89_{-0.55}^{+0.75}$ & $86.92_{-0.24}^{+0.26}$ & photometry \\
\noalign{\smallskip}
$K$ & RV semi-amplitude (m s$^{-1}$) & $93.0\pm2.9$ & $78.7_{-2.8}^{+2.9}$ & radial velocities\\
\noalign{\smallskip}
$e$ & orbital eccentricity & $0.084_{-0.023}^{+0.032}$ & 0 (fixed) & radial velocities\\
\noalign{\smallskip}
$\omega$ & argument of periastron ($^{\circ}$) & $41_{-34}^{+23}$ & \ldots & radial velocities\\
\noalign{\smallskip}
$e\sin\omega$ & & $0.052_{-0.044}^{+0.047}$ & \ldots & radial velocities\\
\noalign{\smallskip}
$e\cos\omega$ & & $0.058_{-0.021}^{+0.019}$ & \ldots & radial velocities\\
\noalign{\smallskip}
$T_{\mathrm{peri}}$ & epoch of periastron (BJD$_\mathrm{TDB}-$2450000) & $7383.43_{-0.30}^{+0.19}$ & \ldots & radial velocities\\
\noalign{\smallskip}
\hline
\noalign{\smallskip}
\multicolumn{5}{l}{\emph{Derived planetary parameters}} \\
\noalign{\smallskip}
$M_P$ & mass ($M_J$) & $0.613_{-0.026}^{+0.027}$ & $0.579_{-0.027}^{+0.028}$ & calculated \\
\noalign{\smallskip}
$R_P$ & radius ($R_J$) & $1.000_{-0.067}^{+0.071}$ & $1.039_{-0.051}^{+0.050}$ & calculated \\
\noalign{\smallskip}
$\rho_P$ & density (g cm$^{-3}$) &  $0.76_{-0.14}^{+0.17}$ & $0.640_{-0.080}^{+0.098}$ & calculated \\
\noalign{\smallskip}
$\log g_P$ & surface gravity (cgs) &  $3.181_{-0.058}^{+0.059}$ & $3.123_{-0.040}^{+0.042}$ & calculated \\
\noalign{\smallskip}
$a$ & semi-major axis (AU) & $0.04097\pm0.00064$ & $0.04839\pm0.00076$ & calculated \\
\noalign{\smallskip}
$T_{\mathrm{eq}}$ & equilibrium temperature (K) &  $1076_{-30}^{+31}$ & $1092_{-20}^{+19}$ & calculated
\enddata
\tablecomments{Stellar and planetary parameters for our targets.}
\tablenotetext{a}{Based on the spectral type vs. effective temperature calibration of \citet{Straizys81}.}
\tablenotetext{b}{$\log g_{\star}$ as measured from stellar spectra.}
\tablenotetext{c}{Micro and macroturbulent velocity are derived using the calibration equation from \citet{Bruntt10} and \citet{Doyle14}, respectively.}
\tablenotetext{d}{TS23 produces iodine cell RVs, which are differential RVs, not absolute like those produced by FIES and HARPS-N. This is the RV offset needed to bring the TS23 data into the same frame as the FIES data, which is {\it not} the same frame as the HARPS-N data. The quoted $\gamma$ values can then be used to bring all three datasets into the same frame.}
\tablenotetext{e}{Limb darkening coefficients are measured in the {\it Kepler} bandpass.}
\tablenotetext{f}{T10 refers to the relations among $T_{\mathrm{eff}}$, $\log g_*$, $[\mathrm{Fe}/\mathrm{H}]$, $M_{\star}$, and $R_{\star}$ found by \cite{Torres10}.}
\tablenotetext{g}{$\log g_{\star}$ as derived from global modeling of the system.}
\tablenotetext{h}{The uncertainties on these isochrone ages are derived solely from the formal uncertainties on the stellar parameters and do not take systematic uncertainties into account. The uncertainties in the ages may thus be severely underestimated.}
\end{deluxetable*}

\section{Conclusions}

We have identified two hot Jupiter candidates in data from K2 Campaign 4, and confirmed the planetary nature of these objects, \planetnine\, and \planetseven, by using radial velocity observations to detect the reflex motion of their host stars. \planetnine\, was discovered independently by \cite{Santerne16}; they additionally gave it a second name, WASP-152\,b, as they also detected it in SuperWASP data. \planetseven\, was discovered independently by both \cite{LilloBox16} and \cite{Brahm16}. All of these works found stellar and planetary parameters which are broadly consistent with ours. Coincidentally, the physical parameters of the planets are very similar to each other; their masses, radii, and equilibrium temperatures (assuming perfect heat redistribution and an albedo of zero) are identical to within $1\sigma$. Both planets have radii typical for their masses and equilibrium temperatures. \planetnine\ orbits a $V=12.56$ mag star, whereas \planetseven\ orbits a relatively faint star with $V=13.53$ mag. Although neither is among the brightest stars to host transiting exoplanets, these stars are nonetheless bright enough to be amenable to ground-based transit observations. We encourage such observations, particularly for \starnine, where the higher spatial resolution of ground-based photometry can easily separate \starnine\, from the nearby contaminating source and thereby eliminate the uncertainties caused by the contamination in the K2 lightcurve. 

Somewhat unusually, the orbit of \planetnine\, is slightly eccentric ($e=0.084_{-0.023}^{+0.032}$), while \starseven\, is slightly metal-poor ($[\mathrm{Fe}/\mathrm{H}]=-0.15 \pm 0.05$). The relatively short rotation period ($10.777 \pm 0.031$ days), high-amplitude rotational variability ($\sim$1-2\%), and presence of chromospheric Ca~\textsc{ii} H \& K emission suggest that \starnine\, may be relatively young, though the lack of detectable Li\,{\sc i} $\lambda$6708~\AA\ absorption line indicates that it is not as young as the $\sim360$ Myr value indicated by gyrochronology.
Additionally, the eccentric orbit of \planetnine\, suggests that either \planetnine\, migrated to its current location via high-eccentricity migration, or that there is an additional planet in the system exciting the eccentricity. These possibilities could be distinguished using long-term RV and/or TTV monitoring to detect an additional companion.

\vspace{12pt}

We thank Nuccio Lanza for assisting with the calculation of the tidal timescales for \planetnine, and Avi Shporer for useful discussions.

This paper includes data taken at The McDonald Observatory of The University of Texas at Austin. Based on observations obtained \emph{a}) with the Nordic Optical Telescope (NOT), operated on the island of La Palma jointly by Denmark, Finland, Iceland, Norway, and Sweden, in the Spanish Observatorio del Roque de los Muchachos (ORM) of the Instituto de Astrofisica de Canarias (IAC); \emph{b}) with the Italian Telescopio Nazionale Galileo (TNG) also operated at the ORM (IAC) on the island of La Palma by the INAF - Fundacion Galileo Galilei. The research leading to these results has received funding from the European Union Seventh Framework Programme (FP7/2013-2016) under grant agreement No. 312430 (OPTICON) and from the NASA K2 Guest Observer Cycle 1 program under grant NNX15AV58G to The University of Texas at Austin. This research has made use of the Exoplanet Orbit Database and the Exoplanet Data Explorer at exoplanets.org.  This publication makes use of data products from the Two Micron All Sky Survey, which is a joint project of the University of Massachusetts and the Infrared Processing and Analysis Center/California Institute of Technology, funded by the National Aeronautics and Space Administration and the National Science Foundation. This publication makes use of data products from the Wide-field Infrared Survey Explorer, which is a joint project of the University of California, Los Angeles, and the Jet Propulsion Laboratory/California Institute of Technology, funded by the National Aeronautics and Space Administration.

\end{document}